\documentclass[preprint,12pt]{elsarticle}
\journal{Computer Physics Communications}
\usepackage[T1]{fontenc} 

\usepackage{url}
\usepackage{graphicx}
\usepackage{axodraw2}
\usepackage{color}
\usepackage{pstricks}

\newcommand{\be}{\begin{equation}}
\newcommand{\ee}{\end{equation}}
\newcommand{\bea}{\begin{eqnarray}}
\newcommand{\eea}{\end{eqnarray}}
\newcommand{\dd}{\mbox{d}}

\newcommand{\al}{\alpha}
\newcommand{\nn}{\nonumber}
\newcommand{\ep}{\varepsilon}
\newcommand{\Gm}{\Gamma}
\newcommand{\I}{i}

\begin{document}

\begin{frontmatter}

\title{
\normalfont
\vskip-1cm{\baselineskip14pt
  \begin{flushleft}
      \normalsize TTP13-047\\
      \normalsize SFB/CPP-13-109
  \end{flushleft}}
  \vskip1.5cm
  FIESTA 3: cluster-parallelizable multiloop numerical calculations in physical regions}

\author[SRCC,KIT]{A.V.~Smirnov\corref{cor1}}
\ead{asmirnov80@gmail.com}

\cortext[cor1]{Corresponding author}

\address[SRCC]{Scientific Research Computing Center, Moscow State University, 119992\\ Moscow, Russia}
\address[KIT]{Institut f\"ur Theoretische Teilchenphysik, Karlsruhe Institute of Technology, D-76128 Karlsruhe, Germany}

\begin{abstract}
The goal of this paper is to present a new major release of the program \\ FIESTA
(Feynman Integral Evaluation by a Sector decomposiTion Approach).
This version presents features like cluster-parallelization, new asymptotic expansion algorithms,
calculations in physical regions,
new sector-decomposition strategies,
 as well as multiple speed, memory, and stability improvements.
\end{abstract}
\begin{keyword}
Feynman diagrams \sep Multiloop Feynman integrals \sep Dimensional regularization \sep Computer algebra
\end{keyword}
\end{frontmatter}
\newpage

{\bf PROGRAM SUMMARY}

\vspace{1cm}

\begin{small}
\noindent
{\em Manuscript Title:} FIESTA 3: cluster-parallelizable multiloop numerical calculations in physical regions\\
{\em Authors:} A.V. Smirnov\\
{\em Program title:} FIESTA 3\\
{\em Licensing provisions:} GPLv2\\
{\em Programming language:} {\tt Wolfram Mathematica} 7.0 or higher, {\tt c++}\\
{\em Computer(s) for which the program has been designed:} from a desktop PC to a supercomputer\\
{\em Operating system(s) for which the program has been designed:} Unix, Linux, \\ Mac OS X\\
{\em RAM required to execute with typical data:} depends on the complexity of the \\ problem  \\
{\em Has the code been vectorised or parallelized?:} yes\\
{\em Number of processors used: } from 1 processor up to loading a supercomputer (tests were performed up to 1024 cores) \\
{\em Supplementary material:} The article, usage instructions in the program package, http://science.sander.su, https://bitbucket.org/fiestaIntegrator/fiesta/overview\\
{\em Keywords:} Feynman diagrams, Multiloop Feynman integrals, Dimensional regularization, Computer algebra\\
{\em CPC Library Classification:} 4.4 Feynman diagrams, 4.12  Other Numerical
Methods, 5 Computer Algebra, 6.5 Software including Parallel Algorithms\\
{\em External routines/libraries used:} {\tt Wolfram Mathematica} [1], {\tt KyotoCabinet} [2], {\tt Cuba} [3], {\tt QHull} [4] \\
{\em Nature of problem:}
The sector decomposition approach to evaluating Feynman integrals
falls apart into the sector decomposition itself, where one
has to minimize the number of sectors; the pole resolution
and epsilon expansion; and the numerical integration of the resulting expression.
Morover, in cases where the integrand is complex, one has to perform a contour deformation\\
{\em Solution method:}
The program has a number of sector decomposition strategies. One of the most important features
is the ability to perform a contour deformation, as well as the so-called preresolution
in case of integrals at threshold.
\\
Everything except the integration is performed in {\tt Wolfram Mathematica} [1] 
(required version is 7.0 or higher). This part of the calculation is parallelizable with the use of shared memory.
The database is stored on hard disk with the use of the {\tt KyotoCabinet} [2] database engine.
\\
The integration part of the algorithm can be performed on a cluster, is written in {\tt c++} and does not need 
{\tt Wolfram Mathematica}. For integration we use the Cuba library package [3].
\\
{\em Restrictions:} The complexity of the problem is mostly restricted
by CPU time required to perform the integration and obtain a proper precision.\\
{\em Running time:} depends on the complexity of the problem.\\
{\em References:} 
{\\} [1] http://www.wolfram.com/mathematica/, commercial algebraic software; 
{\\} [2] http://fallabs.com/kyotocabinet/, open source; 
{\\} [3] http://www.feynarts.de/cuba/, open source; 
{\\} [4] http://www.qhull.org, open source.

\end{small}

\newpage

\section{Introduction}

Let us consider a Feynman integral

\bea
  \mathcal F(a_1,\ldots,a_n) &=&
  \int \cdots \int \frac{\dd^d k_1\ldots \dd^d k_h}
  {E_1^{a_1}\ldots E_n^{a_n}}\,,
  \label{eqbn-intr}
\eea
where the denominator factors $E_i$ are linear functions with respect to
scalar products of loop momenta $k_i$ and external momenta $p_i$, and dimensional regularization with
$d=4-2\epsilon$ is implied.

Such an integral depends on masses and kinematic invariants --- scalar products of external moments.
After substituting all values for kinematic invariants and masses one can evaluate the integral numerically.
This can be done automatically with the so-called sector-decomposition approach, 
initially introduced by Binoth and Heinrich~\cite{Binoth:2000ps,Binoth:2003ak,Binoth:2004jv,Heinrich:2008si,Bogner:2007cr,Bogner:2008ry}.

This approach is based on the so-called alpha-representation of Feynman integrals:

\begin{eqnarray}\label{Alpha}
    &&\mathcal F(a_1,\ldots,a_n) =(i\pi^{d/2})^l\times
   \\\nonumber
    &&\frac{\Gamma(A-l d/2)}{\prod_{j=1}^n \Gamma(a_j)}
            \int_{x_j\geq 0} d x_i\ldots d x_{n} \delta\left(1-\sum_{i=1}^n x_i \right)
                \left(\prod_{j=1}^n x_j^{a_j-1}\right) \frac{U^{A-(l+1)d/2}}{F^{A-ld/2}},                         
\end{eqnarray}
where $A=\sum_{i=1}^n a_n$, $l$ is the number of loops and
$U$ and $F$ are constructively defined polynomials of $x_i$.

There are three public programs that can perform the numerical calculation with the sector decomposition
approach --- {\tt sector\_decomposition} by Bogner and Weinzierl~\cite{Bogner:2007cr,Bogner:2008ry}, {\tt SecDec} by Binoth and Heinrich~\cite{Heinrich:2008si} 
(later improved and made public by Borowka, Carter and Heinrich~\cite{Carter:2010hi,Borowka:2012yc,Borowka:2012ii,Borowka:2012rt,Borowka:2013cma,Borowka:2013lda}) and {\tt FIESTA}~\cite{Smirnov:2008py,Smirnov:2009pb} initially created by the author of this paper with M.~Tentukov and later improved together with V.~Smirnov. In this paper we present 
a new version of {\tt FIESTA}.

Both {\tt FIESTA} and {\tt FIESTA 2} had a major disadvantage when being compared with {\tt SecDec 2} --- they were not able
to perform calculations in physical regions. Hence some of the advantages such as the internal code compiler 
designed especially for sector integrals and the multi-precision calculation could not be used by those
interested in physical regions. The new version fills this gap by utilizing the ideas 
initially suggested in \cite{Soper:1999xk,Kurihara:2005ja,Binoth:2005ff,Nagy:2006xy,Anastasiou:2007qb}
and presented as an algorithm in the second version of {\tt SecDec}~\cite{Carter:2010hi}.

One more important new feature is the ability to use cluster parallelization. If one is going to perform multiloop calculations and aims at
high-precision results, it may take a lot of time to obtain those. The new version of {\tt FIESTA} has an improved internal structure
so that the algebraic part can be performed by {\tt Wolfram Mathematica} on a single computer and the results
can be saved into a database. Afterwards the integration can be performed on a cluster with the use of {\tt MPI}-parallelization.
This approach also helps to deal with the {\tt Mathematica} licensing policy --- the integration part does not need {\tt Mathematica}
licenses anymore.

Although the sector decomposition approach is quite powerful, 
in some cases the integral is too complicated for direct evaluation. Even if the calculation of master integrals
can be performed, the other important part, IBP reduction~\cite{Chetyrkin:1981qh} might fail. In this case one can use the asymptotic expansion approach. 
The kinematic invariants are separated into groups proportional to different groups
of a small parameter $\rho$. One is interested in the behavior of Feynman integrals
when $\rho$ tends to zero.

There are different approaches to asymptotic expansion of Feynman integrals,
but we are interested in numerical algorithms. The reason is that those algorithms
can be implemented as computer codes so that the asymptotic expansion can be done
completely automatically. 

In this paper we consider two numerical algorithms of asymptotic expansion.
Both of them use sector decomposition and hence become a part of the new version of {\tt FIESTA}.

The first algorithm uses a Mellin-Barnes integration combined with sector decomposition ---
it has already been encoded in FIESTA2~\cite{Smirnov:2009pb}. The disadvantage of this algorithm is that
it can be applied only in situations where the kinematic invariants can be separated into
two groups --- large and small. Moreover this algorithm encounters some
computer problems --- the complexity of evaluations grows high enough to some make
three-loop problems too complicated for this approach. 

The second algorithm is proposed in this paper. It consists of applying the {\tt asy} code~\cite{Pak:2010pt,Jantzen:2012mw}
in order to reveal regions~\cite{Beneke:1997zp,Smirnov:1999bza}. Then the algorithm 
produces contributions of regions by expanding the integrand in a certain way
and applies the sector decomposition in order to calculate expansion coefficients. 
However in many situations the corresponding integrals happen not to be well-defined ---
one needs to add an extra regularization in order to calculate them. 
This makes the sector decomposition approach more complex.

In section 2 we explain how {\tt FIESTA} treats integrals at and above the threshold, in section 3 we describe
the stages that {\tt FIESTA} performs and the parallelization during those stages. 
In section 4 we present two algorithms used for numerical asymptotic expansion and in section 5 we provide
installation and usage instructions.

\section{Integrating at and above the threshold}

\label{complex}

Sector decomposition strategies are guaranteed to terminate while the function $F$ from formula~(\ref{eqbn-intr}) has no monomials with negative coefficients.
If due to some values of kinematic invariants $F$ has negative terms, this approach may fail. 
In principle, there are three different variants:

\begin{enumerate}
 \item Below threshold. Some terms of $F$ are negative, but it still turns to zero only due to some variables $x_i$ turning to zero. In this case the sector decomposition works as normal;
 \item At threshold. $F$ is never negative, but might turn to zero at some internal points. To solve this problem we use the so-called pre-resolution approach suggested in {\tt FIESTA 2}\cite{Smirnov:2009pb} and improved in {\tt FIESTA 3} (for details see \cite{Jantzen:2012mw}). Some transformations are performed before the sector decomposition aimed to get rid of singularities of the form $(x_i-x_j)^2$. We have multiple confirmations that this approach works well for Feynman integrals;
 \item Above threshold. Here $F$ can be of different signs in different parts of the integration domain and, therefore turns to zero at multiple points inside. The integral is a complex number, however the most complicated part is that the integration cannot be taken directly due to the zero values of $F$. An automatic way to deal with this problem was suggested by Binoth and Heinrich~\cite{Binoth:2000ps,Binoth:2003ak} and implemented by Borowka and Heinrich in the second version of {\tt SecDec}~\cite{Borowka:2012rt}. The basic idea is to transform the contour in order to shift away from the zero values of $F$. This makes the integration process slower but provides an automatic way to process integrals above threshold. We also implemented this strategy in {\tt FIESTA 3} mostly following the guidelines from~\cite{Borowka:2012rt}. In order to activate this algorithm one has to set {\tt ComplexMode=True} in {\tt FIESTA 3}.
\end{enumerate}

The contour transformation algorithm described in~\cite{Borowka:2012rt} and used after a modification in {\tt FIESTA} consists of the following steps
(each step uses {\tt LambdaIterations} iterations):
\begin{enumerate}
 \item The maximal values of shift functions $S_i := x_i (1 - x_i) \frac{d F}{d x_i}$ are estimated (we substitute a number of random points); if a maximum is bigger than $1$, the corresponding shift function is multiplied by its inverse;
 \item The integration variables $x_i$ are replaced with $x_i - \lambda \mathrm{i} S_i$, the dependence of $F$ on the new variables is considered;
 \item The code estimates the maximal possible value $\lambda$ such that the cubic terms of $F$ do not exceed the linear terms. This is close to checking that the function $F$ does not change the sign of its complex part (as specified by the dimension-regularization prescriptions). If the resulting value is bigger that {\tt MaxComplexShift} the latter is used;
 \item The code splits the interval from $0$ to the found maximal possible $\lambda$ into the number of parts equal to {\tt LambdaSplit}. For each of those values a new check is performed and a best value is chosen: we require that the sign of the complex part of $F$ is always negative, and among those choices find such $\lambda$ that the minimal absolute value of $F$ is maximal;
 \item The $\lambda$ value is fixed and substituted into the deformation formula. 
\end{enumerate}

Those calculations are performed independently for each sector. 

Alternatively, one can specify the value of {\tt ComplexShift} to set a fixed $\lambda$ and turn off this algorithm.
Since the search in {\tt Mathematica} for best possible $\lambda$ is performed randomly and the speed of {\tt Mathematica} calculations is not too fast,
one cannot be absolutely sure that $\lambda$ is chosen properly --- the sign of the imaginary part of $F$ might turn positive, and one obtains a wrong answer.
It is even more risky if one chooses a fixed complex shift. Hence it is recommended to perform extra sign tests in the {\tt c++} part of the program.

To do so one should run the code with the {\tt OnlyPrepare} option so that the integrands are stored in a database.
After that {\tt Mathematica} prints a command to integrate everything, but one should first add the {\tt -testF} argument to this command (see section \ref{PoolOptions}).
One should ignore the answer produced by the integrator and only watch whether it succeeds. It means that $\lambda$ is chosen well and one can turn back to the integration.

The contour deformation does not help if $F$ turns to $0$ due to some variables being equal to $1$. If one suspect that this might be the case, the {\tt TestF1=True} setting should be used while running {\tt FIESTA}. Then one will be informed on variables leading to singularities of this sort. Then one should use the {\tt BisectionVariables} setting listing the problematic variables. This makes {\tt FIESTA} break the integration into two parts --- from $0$ to $p$ and from $p$ to $1$ for some $p$. Then in the first part we make the variable replacement $x=y p$ and in the second: $x=1-y (p-1)$. Thus the singularities at $x\rightarrow 1$ are changed to $x\rightarrow 0$ and are treated by the standard sector decomposition. By default the middle point $p$ is equal to $1/2$, but a different value can be chosen by the {\tt BisectionPoint} setting or different values for different variables by the {\tt BisectionPoints} list.

\subsection{Example}

\begin{figure}[!htb]
\centering
\begin{picture}(150,60)(0,0)

\Vertex(50,0){1.5}
\Vertex(50,50){1.5}
\Vertex(100,0){1.5}
\Vertex(100,50){1.5}
\Line(35,0)(115,0)
\Line(35,50)(115,50)
\Line(50,0)(50,50)
\Line(100,0)(100,50)
\Line(50,0)(100,50)
\Line(100,0)(50,50)
\Text(28,0)[]{$p_2$}
\Text(124,0)[]{$p_4$}
\Text(28,50)[]{$p_1$}
\Text(124,50)[]{$p_3$}
\end{picture}

\caption{K4 graph}

\label{K4}
\end{figure}
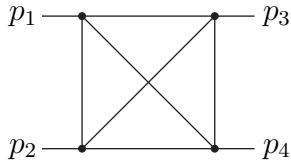

Let us consider an example (see fig.\ref{K4}) --- an on-shell massless K4-graph (we choose $s=1$ and $t=-1/4$). The answer is a complex number, so the contour transformation is used. 

The analytic answer for the corresponding integral was recently obtained in~\cite{Henn:2013nsa}; the coefficient at $\ep^2$ 
is equal to $1024.2413 + 889.3892 i$ (we cut it to $8$ digits), so we can compare the numerical results with it.
We performed the calculations on a computer having two Hexa-Core Intel X5675 3.07 GHz processors. The database preparation step took less than 10 minutes.
The integration time and the numerical uncertainties depend on the number of sampling points and are presented in Table~\ref{K4table}. The columns contain the number of sampling points, time needed to calculate the $\ep^2$ part, the numerical coefficient at $\ep^2$, the error estimate and the real error. The default setting for the number of sampling points is $50 000$.

\begin{table}[htb]
\begin{tabular}{c|c|c|c|c}
Points   & Time (sec.) & Result & Error est.(\%) & Error (\%) \\\hline
5 000    &    $224$  &  $1023.64 + 889.577 i$ & $0.0508$ &  $0.0465$  \\\hline
50 000    &    $3339$  &  $1024.33 + 889.412 i$ & $0.0183$ &  $0.0068$  \\\hline
500 000     &     $42940$  &  $1024.29 + 889.387 i$ & $0.0058$ & $0.0035$  \\\hline
5 000 000     &     $426726$ &  $1024.25 + 889.388 i$ & $0.0018$ &  $0.0007$ 
\end{tabular}
\caption{Results and timings for the K4 graph}
\label{K4table}
\end{table}

We can see from table~\ref{K4table} that both the error estimate and the real error are decreasing proportional to the square root of the number of sampling points.
The time grows linearly at large scales. The less than linear growth for smaller numbers can be explained by the fact that some small calculations might be performed with the use of processor cache and hence faster, but starting with some point the cache is not enough and the dependence becomes linear. The result is always within the error estimate, although we have to state that the error estimate is produce by the {\tt Vegas} algorithm and is only a probability prediction but not a guarantee. 

\section{Cluster-parallelization and internal structure of FIESTA 3}
Let us describe the internal structure of {\tt FIESTA 3}.
The input for {\tt FIESTA} is the list of propagators, the list of internal moments,
the list of indices and the requested order of epsilon.

The algorithm consists of three major stages. The \textbf{first stage} is the initial preparation
and sector decomposition. The algorithm performs the following:

\begin{itemize}
 \item Eliminating negative indices. If some of the indices are non-positive, the algorithm 
 differentiates the integrand by the corresponding alpha-parameters according to the following rule:
\[
\int_{0}^{\infty}d x \frac{x^{(a-1)}}{\Gamma(a)} f(x) = f^{(n)} (0)
\]
for non-positive integer $a$.

 \item ``Pre-resolution''. If we consider an integral on a threshold, the function $F$ might have negative terms, 
 but is non-negative totally. However it might be equal to zero somewhere in the middle in the integration domain.
 This might be a problem for the integration, hence we perform the so-called pre-resolution. The algorithm searches 
 for pairs of variables that might produce negative terms due to contributions like $(x_i-a x_j)^2$, divides the infinite
 sector into two and makes appropriate variable replacements. The details can be found in~\cite{Jantzen:2012mw}.
 
 \item Sector decomposition. A number of strategies can be used including the strategy by Binoth and Heinrich, 
 by Bogner and Weinzierl, the original strategy from {\tt FIESTA 1}, the Hepp and Speer sector strategies~\cite{Smirnov:2008aw},
 as well as the most effective (by the number of sectors) but relatively slow strategy by Kaneko and Ueda~\cite{Kaneko:2009qx}.
 
 This part is parallelized in {\tt Mathematica} --- different kernels work on different primary sectors.
 
\end{itemize}

After the first stage is over, a database is prepared. We use the kyotocabinet (\url{http://fallabs.com/kyotocabinet/}) engine for storing data.
Now each sector can be treated independently, hence we use the {\tt Mathematica} parallelization in order to speedup the calculations.
The \textbf{second stage} consists of a number of parts. Each time the expressions are read from one database
and written to the other one. Only the main {\tt Mathematica} process accesses the databases, the subkernels
only perform equivalent tasks in different sectors. There are the following parts:

\begin{itemize}
 \item Variable substitution. The sector decomposition variable replacements are performed (the replacements rules were created at the previous stage). 
 \item Contour transformation (only in complex mode). The contour is shifted to avoid the zeros of $F$.
 \item Pole resolution. If the integrer parts of some exponents of sector variables are non-positive, then in order to reveal singularities the integrand is treated as the first few terms of Taylor series plus the remainder.
 \item Expression preparation. All functions are combined and multiplied.
 \item Epsilon expansion. The integrands can now be expanded in epsilon up to the required degree.
 \item String preparation. The input for the {\tt c++} code is prepared. Now the expressions in the database are no longer {\tt Mathematica} expressions but strings ready to be taken by the integrator.
\end{itemize}

If the user specifies the {\tt OnlyPrepare} mode, then at this point the algorithm stops. As a result one has a database with integration strings.
Starting from this point, one can run the integration without {\tt Mathematica}. However if {\tt OnlyPrepare} is set to {\tt False}, {\tt Mathematica}
runs an integrator itself and waits for results.

The integration part has the following structure. There are two possible binaries that can be used --- {\tt CIntegratePool} and {\tt CIntegratePoolMPI}.
The second one is intended for cluster usage with the {\tt MPI}-parallelization, the first one uses threads. 
None of these binaries performs the integration itself. It only launches the real integrator processes, {\tt CIntegrate} (basic variant),
{\tt CIntegrateMP} (variant with {\tt MPFR}) or {\tt CIntegrateMPC} (complex variant with {\tt MPFR}). The threads version simply runs a requested number of threads, each of those forks and starts the integrator.
So the main process reads from the database and distributes tasks between slaves and gathers results afterwards.
If the {\tt MPI} version is used, only one copy accesses the database as well, the tasks to other nodes are send via {\tt MPI}.

We use the integrators from the Cuba library~\cite{Hahn:2004fe} by Hahn, however the evaluation of functions is performed in an original way.
Unlike it is done in other sector decomposition programs, we do not compile the {\tt c++} integrand by means of standard compilers.
On the contrary, the expression is analyzed inside the {\tt c++} part of {\tt FIESTA} and is transformed into an optimized internal form allowing
fast numerical function evaluation. Moreover, the expression is analyzed in order to decide, where to use double precision and where it
is important to use the multi-precision evaluations (for details see~\cite{Smirnov:2009pb}).

We prepared a database and then tested how the {\tt MPI}-parallelization works for a sample massive on-shell four-loop propagator diagram
(with propagators $\{-(l_1 + q)^2 , -(l_2 + q)^2 , -(l_3 + q)^2 
        , -(l_4 + q)^2 , -(l_4)^2 + M^2, -(l_2 - l_3 + l_4)^2+         M^2, -(l_1)^2 + M^2, -(l_1 - l_2)^2 , -(l_4 - l_3)^2
                , -(l_1 - l_2 + l_3 - l_4)^2 , -(l_2 - l_3)^2 
                    \}$ where $M^2=q^2$), the results are shown in Table~\ref{timings}.

\begin{table}[htb]
\begin{tabular}{c|c|c|c|c|c|c}
Points $\backslash$ Cores  & 32 & 64 & 128 & 256 & 512 & 1024 \\\hline
500 000          & 5112 & 2643 & 1570 & 1381 & 1344 & 1330 \\\hline
5 000 000          & 48305 & 24719 & 13524 & 7833 & 5116 & 3795 \\\hline
50 000 000          & 482948 & 245109 & 133917 & 77951 & 50734 & 36988
\end{tabular}
\caption{Timings of a 4-loop massive propagator}
\label{timings}
\end{table}

The columns of the table show the number of cores used during the parallelization, the rows --- the number of sampling integration points used during the evaluation. The result is measured in seconds. 
This table indicates that massive parallelization is not very effective if the number points is not too high --- this happens due to the database and {\tt MPI} overhead. However this part of the evaluation has a fixed time and while the number of sampling points is increased, the efficiency of the parallelization grows well. 

As for the precision obtained, it grows approximately as the square root of the number of sampling points used. Hence the {\tt MPI}-parallelization can be a good way to improve  results, however this is possible only if one has enough CPUs times hours. For example, the evaluation with 50 million sampling points with 1024 cores (the tests were performed on the ``Lomonosov'' supercomputer~\cite{opanasenko2013lomonosov3827487}; the Intel Xeon 5570 Nehalem 2.93 GHz processors connected with QDR InfiniBand, 40 Gbit/sek were used) took 100 hours resulting in a rather big number about 10 thousand CPUs times hours. Possibly it cannot grow much further because the number of jobs in this test was equal to 11161, and the number of cores should be a lot less than the number of cores in order to obtain an efficient parallelization.

A similar table was created with the {\tt SeparateTerms} setting, when the integators are split more than one integrand per sector (resulting in 35971 {\tt MPI} jobs instead of 11161). The results are displayed in Table~\ref{timings2}. We can see a better dependence of integration time on the number of cores in this case.

\begin{table}[htb]
\begin{tabular}{c|c|c|c|c|c|c}
Points $\backslash$ Cores  & 32 & 64 & 128 & 256 & 512 & 1024 \\\hline
500 000          & 4684 & 2345 & 1193 & 614 & 452 & 451 \\\hline
5 000 000          & 41596 & 20720 & 10431 & 5355 & 2826 & 1579 \\\hline
50 000 000          & 399997 & 200004 & 100555 & 51467 & 27149 & 15195
\end{tabular}
\caption{Timings with the {\tt SeparateTerms} setting}
\label{timings2}
\end{table}

\begin{figure}[htb]
\includegraphics[width=\textwidth]{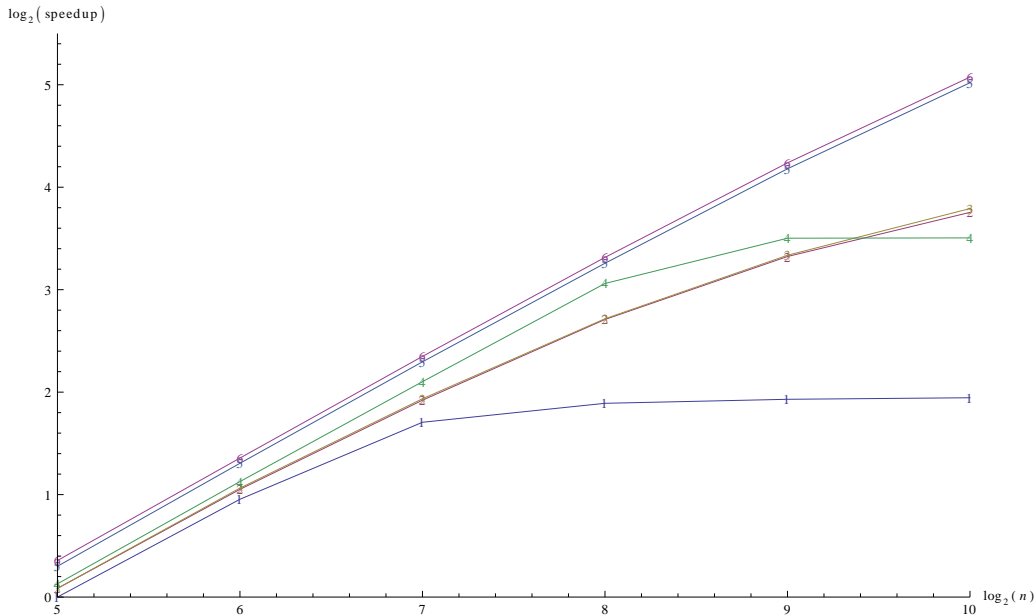}
\label{timings3}
\caption{Graphical representation of timings}
\end{figure}

These numbers are also presented in Diagram~\ref{timings3}. As a base value we take $time(32,50000)$ --- the time in seconds needed to evaluate the integral with $32$ cores and $50000$ sampling points.
The $y$-axe has $log_2(speedup)$, where $speedup = (time(n,points)/ time(32,50000)) * (50000/points)$ (the second factor is here because we measure the speedup by the number of cores).
The functions $1-3$ on the diagram correspond to the dependences $time(n,50000)$, $time(n,500000)$ and $time(n,500000)$, the functions $4-6$ --- are the same but with the {\tt SeparateTerms} setting. We can clearly see that with the {\tt SeparateTerms} setting and a large enough number of sampling points the integration scales really well and suits perfectly to be run on large clusters.

\section{Numerical algorithms for asymptotic expansion of Feynman integrals}

\subsection{Mellin-Barnes and sector decomposition}

This approach, initially suggested by Pilipp~\cite{Pilipp:2008ef}, is based on combining the Mellin-Barnes and
the sector decomposition approaches. 
Let us write the integral in its alpha-representation (for explicit coefficient values see formula~\ref{Alpha}):

\begin{eqnarray}
\label{eqn:alprep}
  &&\mathcal F(a_1,...,a_n) =  \\ \nonumber && c \int_0^\infty d x_1 ... d x_n~
  \delta(1-x_1-...-x_n) x_1^{a_1-1}...x_n^{a_n-1}
  U^a F^b,
\end{eqnarray}

Suppose that the function $U$ does not depend on kinematic invariants and $F$ can be written as
$W_1+W_2$, where the terms in $W_1$ are much smaller that the terms in $W_2$.
Let us introduce a small parameter $\rho$ and replace $W_1$ with $\rho W_1$
(now $\rho$ is small and terms of $W_1$ are not).
The parts of $F$ now can be separated by introducing a single Mellin-Barnes integration:

\be
\frac{\Gm(a-h d/2)}{(\rho W_1+W_2)^{a-h d/2}}= \frac{1}{2\pi
i}\int_{-i \infty}^{+i \infty} \dd z\, \rho^z \frac{\Gm(a-h
d/2+z) \Gm(-z)}{W_1^{-z} W_2^{a-h d/2+z}}  \; \, \label{MB1} \ee
so that we obtain 

\bea &&{\cal F}(a_1,...,a_n)  =\frac{\left(\I\pi^{d/2}
\right)^h}{\prod_l\Gm(a_l)} \frac{1}{2\pi i}\int_{-i \infty}^{+i
\infty} \dd z\, \Gm(a-h d/2+z) \Gm(-z) \rho^z
\\\nn  && 
\times\int_0^1 \ldots\int_0^1 \hat{\cal U}^{a-(h+1) d/2} \,
W_1^{z} \, W_2^{-a+h d/2-z}\, 
\delta(1-x_1-...-x_n) x_1^{a_1-1}...x_n^{a_n-1}
\;. \label{primsec1} \eea

Now the integration has normal sector integrals (from $0$ to $1$ after taking out the $\delta$ function) and
and extra integration (over variable $z$) from $-i\infty$ to $+i\infty$. 
The exponents depend on the integration variable $z$.

The algorithm builds sectors as if there was no extra variable $z$. Now it considers the variables $\al_i$
with corresponding powers depending on $z$. The integration contour can be closed to the right, but
the first few poles have to be considered (the number of the poles depends on the coefficient at $z$).

Afterwards the algorithm reveals singularities in $\ep$ generated by
the MB integration over $z$. The integral of $t_i^{b_i \ep + c_i z
+n_i-1}$ generates a $z$-dependence of the type $\Gamma\left(b_i \ep
+ c_i z +n_i \right)$. We are concentrating on sector integrals
with $c_i<0$ because they are relevant to our limit.

After revealing those singularities the algorithm can return to normal steps
used in numerical sector decomposition --- expansion by $\ep$ and numerical integration
by a Vegas integrator.

\subsection{Regions and sector decomposition}

Another approach presented in this paper also uses sector decomposition but now together with
the regions approach. The method of regions~\cite{Smirnov:1999gc,Tausk:1999vh}
defines prescriptions to find regions, or
scalings of momentum components that after the expansion
provide non-zero contributions. In each region, we first
Taylor expand the integrand and drop the scaling restrictions.

In \cite{Pak:2010pt} we presented the regions approach written in terms of alpha-re\-pre\-sen\-tation.
It looks natural to make an algorithm that considers those regions; in each region we can expand the integrand 
and evaluate the corresponding integrals numerically afterwards.

This idea encounters a number of technical difficulties, but what is most important, the integrals in regions might be not well defined.
One needs to introduce an extra regularization parameter $\lambda$
(the poles in $\lambda$ get canceled out after summing over all regions).
We make the indices (powers of alpha-parameters) depend linearly on $\lambda$.
This does not affect the determination of regions.

After finding the regions, the algorithm expands the integrand in each of them up to the required degree.
Then it starts the evaluation of expansion coefficients. Those functions no longer depend
on the small parameter $\rho$ but have an extra dependence on the small parameter $\lambda$.

Now the sector decomposition is performed. In each sector we get an integral over a unit hypercube,
where the exponents depend on $\ep$ and on $\lambda$.
The algorithm considers the exponents depending on $\lambda$ plus some negative integer
and performs a ``pre-resolution'' --- replaces the integrand with the few terms of Taylor expansion in $\lambda$
plus the remainder (such that the corresponding integral has no poles over $\lambda$). Afterwards the integrands
can be expanded in $\lambda$ up to order $0$.

Now the algorithm proceeds with standard sector decomposition steps keeping in memory that different terms
are proportional to different integer powers of $\lambda$.
Finally all the coefficients of the expansion in $\lambda$ and $\ep$ are evaluated numerically.

After calculating the result in all regions, the algorithm sums them up and gathers the coefficients. At this point the poles in $\lambda$ should 
cancel, however, they can exist due to numerical uncertainties.

There are some more ideas that improve the algorithm. The expanded region integrals often are more complicated than
the initial ones, so it might be problematic for the numerical integration. However, a few integrations can be taken
out analytically. Quite often after the expansion one of the functions $U$ and $F$ no longer depend on one of the variables
$x_i$. In this case the integration over $x_i$ is taken out analytically.

\section{Installation and usage}

\subsection{Installation}

In order to install {\tt FIESTA 3} one has to download the package from the following address: 
\url{http://science.sander.su/FIESTA.htm}. Then one should compile the c++ part of the code.
Although {\tt FIESTA 3} can be used without the {\tt c++} part (by setting {\tt UsingC=False} and {\tt UsingQLink=False}),
it is not recommended since one will not have access to most features existing in this program.

Hence first one needs to compile {\tt FIESTA}. In order to do that, the following libraries have to exist on the computer:

\begin{itemize}
 \item {\tt KyotoCabinet} --- a fast database engine. It can be downloaded from \url{http://fallabs.com/kyotocabinet/}. Basically it is compiled and installed with
 {\tt \newline ./configure \newline make \newline make install \newline}
 If one cannot install it system-wide, one is able to provide the paths to {\tt KyotoCabinet} later. 
 If one is aiming at a static version of {\tt FIESTA}, it should be configured with {\tt --enable-static --disable-shared}.
 
 \item {\tt Cuba} integrator library by Thomas Hahn \cite{Hahn:2004fe} that can be downloaded from \url{http://www.feynarts.de/cuba/}. We recommend to build it with
  {\tt \newline ./configure \newline make lib \newline make install \newline}
  If one runs {\tt make} instead it might require extra dependencies.
  
 \item {\tt MPFR} --- multiple-precision floating-point library that can be downloaded from \url{http://www.mpfr.org/} but has a big chance to exist in the system repositories. If one is downloading it from the official web-page and is not installing it system-wide, it is recommended to configure it with a prefix and install into a local directory. 
 
 \item {\tt GMP} --- GNU multi precision library, it is also normally installed system-wide. If not, one can download it from \url{http://gmplib.org/}, configure and make. This library is not required at the object compilation step, and for linking one might point to the {\tt .libs} directory in the {\tt GMP} folder.
 
 \item In case one is going to use strategies {\tt KU}, {\tt KU0}, {\tt KU2} or the {\tt SDExpandAsy} command, then one needs to have the {\tt qhull} convex hull search package installed. It might exist in the system repositories or can be downloaded from \url{http://www.qhull.org/}.
 
 \item If one is going to build an {\tt MPI} version of {\tt CIntegratePool}, then one should need to have one of the {\tt MPI} environments to be installed on the system. Normally this is recommended if one is installing {\tt FIESTA} on a cluster.
 
\end{itemize}

Moreover, if one is going to perform both the database preparation and the integration, it is needed to have {\tt Wolfram {\tt Mathematica} 7.0} or higher installed on the computer. However, if the databases with integrands
are going to be prepared elsewhere, and one is only running the integration (a typical case for clusters), {\tt Mathematica} is not required.

Now one can try to build {\tt FIESTA}. If the libraries mentioned earlier are installed and exist on the system paths, then one should simply {\tt cd} to the directory with {\tt FIESTA} and run the {\tt make} command. If not, one should first edit {\tt paths.inc} and add those paths with -I and -L for include and link paths correspondingly.

In rare cases one needs also to edit {\tt libs.inc} to tune library linking.
If one wants it static as much as possible, then {\tt libs.inc} should be replaced with {\tt libs-static.inc}
Normally one does not need to edit the Makefile, neither the main one, nor the ones in subfolders.

The {\tt make} command should build everything but the {\tt MPI} version of CIntegratePool.
The {\tt MPI} version is build with {\tt make mpi} or together with other packages with {\tt make all}.
There is no {\tt make install} in the package.

In order to build {\tt KLink} one needs to have {\tt Mathematica} installed on the system. The paths should be detected automatically.
If one does not want to build {\tt KLink} (the database will be prepared elsewhere) and wishes to avoid compilation errors, {\tt make noklink} should be used.

The build results might be verified with {\tt make test} or {\tt make testall} (to test also {\tt MPI}). No errors should be produced.
These simple tests mainly test whether the binaries are functional.

\subsection{Program syntax}

Let us start with basic examples. If one is loading {\tt FIESTA} from {\tt Ma\-the\-ma\-tica}, it should either be loaded with
\newline
{\tt SetDirectory[<path to FIESTA>]; Get["FIESTA3.m"];} 
\newline
or 
\newline
{\tt FIESTAPath=<path to FIESTA>; Get[FIESTAPath<>"FIESTA3.m"];}.
\newline
Do not load {\tt FIESTA} by simply specifying a full path to it, it will not work properly.

Now one can call the following commands:

{\tt SDEvaluate[\{U,F,l\},indices,order]},

where $U$ and $F$ are the functions from eq.~(\ref{Alpha}), $l$ is the number of loops,
$indices$ is the set of indices and $order$ is the required order of $\varepsilon$-expansion.

To avoid manual construction of $U$ and $F$ one can use a build-in function {\tt UF}
and launch the evaluation as

{\tt SDEvaluate[UF[loop\_momenta,propagators,subst],indices,order]},

where $subst$ is a set of substitutions for external momenta, masses and
other values (to remind: the code performs numerical integration so the functions $U$ and $F$
should not depend on anything external).

\textit{Example}:  

{\tt SDEvaluate[UF[\{k\},\{-k$^2$,-(k+p$_1$)$^2$,-(k+p$_1$+p$_2$)$^2$,-(k+p$_1$+p$_2$+p$_4$)$^2$\},
\\
\{p$_1^2\rightarrow$0,p$_2^2\rightarrow$0,p$_4^2\rightarrow$0,
p$_1$ p$_2\rightarrow$-S/2,p$_2$ p$_4\rightarrow$-T/2,p$_1$ p$_4\rightarrow$(S+T)/2,
\\
S$\rightarrow$3,T$\rightarrow1$\}],
\{1,1,1,1\},0]
}

performs the evaluation of the massless on-shell box diagram.


In the following commands we will only provide the first version of the syntax (with {\tt \{U,F,l\}}). However, in all places this triple can be replaced with the {\tt UF} generator.

Now to expand a Feynman integral by a small parameter {\tt tt} one should use

{\tt SDExpand[\{U,F,l\},indices,order,tt,order in tt]}.

\textit{Example}:  

{\tt SDExpand[UF[\{k,
   l\}, \{-k$^2$, -(k + p1)$^2$, -(k + p1 + p2)$^2$, -l$^2$, \newline
   -(l - k)$^2$,  -(l + p1 + p2)$^2$, -(l + p1 + p2 + p4)$^2$\}, \{p1$^2$ -> 0, p2$^2$ -> 0,
             p4$^2$ -> 0, p1*p2 -> s/2, p2*p4 -> t/2, p1*p4 -> -(s + t)/2,
                s -> -1, t -> -tt\}], \{1, 1, 1, 1, 1, 1, 1\} , 0, tt, 0]}

The new algorithm presented in this paper can also be used to expand this integral, however one has to provide
a regularization variable with the {\tt RegVar} setting and shift indices.

\textit{Example}:  

\tt
RegVar=la;

SDExpandAsy[UF[\{k,
   l\}, \{-k$^2$, -(k + p1)$^2$, -(k + p1 + p2)$^2$, -l$^2$, -(l - k)$^2$, -(l +
          p1 + p2)$^2$, -(l + p1 + p2 + p4)$^2$\}, \{p1$^2$ -> 0, p2$^2$ -> 0,
             p4$^2$ -> 0, p1*p2 -> s/2, p2*p4 -> t/2, p1*p4 -> -(s + t)/2,
                s -> -1, t -> -tt\}], \{1+la, 1, 1, 1-la, 1, 1, 1\} , 0, tt, 0]

\normalfont

To use the Speer sectors strategy one should use {\tt SDEvaluateG} instead of {\tt SDEvaluate}.
The syntax is

{\tt SDEvaluateG[graph\_info,\{U,F,l\},indices,order]} 

or 

{\tt SDExpandG[graph\_info,\{U,F,l\},indices,order,tt,tt\_order]}.

The graph information should be of the form $\{lines,$ $external\_vertices\}$,
where $lines$ is a list of pairs of vertices connected by this line. The
vertices should be numbered from $1$ without skipping numbers. It is also very
important to have the order of lines coincide with the order of propagators in
the $UF$ input. For example, for the tetrahedral input is the following.

\textit{Example}:  

\tt

SDEvaluateG[\{\{\{1, 2\}, \{2, 3\}, \{3, 1\}, \{4, 2\}, \{4, 3\}, \{4, 1\},

UF[\{k1, k2, k3\},
\{-k1$^2$ + 1, -k2$^2$ + 1, -k3$^2$ + 1, 

-(k1 - k2)$^2$ + 1, -(k2 - k3)$^2$ + 1, -(k3 - k4)$^2$ + 1,
\},
 \{\}],

\{1, 1, 1, 1, 1, 1, 1, 1\}, -1]

\normalfont

There is one command that makes possible to apply sectors decomposition to integrals different from Feynman integrals:

{\tt SDEvaluateDirect[var,function,degrees,order,deltas\_optional]}.

Here {\tt var} stands for the integration variable used in functions (for example, {\tt x} goes for {\tt x[1]}, {\tt x[2]}, ...),
{\tt functions} is a list of functions and {\tt degrees} is the list of their exponents. {\tt order} is the requested order is epsilon,
and {\tt deltas} goes for the list of delta functions attached to the integrand. By default is is empty. For example,
{\tt \{\{1,3\},\{2,4\}\}} goes for the product of {\tt Delta[x[1]+x[3]-1]} and {\tt Delta[x[2]+x[4]-1]}.

\textit{Example}:  

\tt
SDEvaluateDirect[x, \{1, 
  x[1] x[2] x[3] + x[1] x[2] x[4] + 
  
  x[1] x[3] x[4] + x[1] x[2] x[5], 
  x[1] x[3] + x[2] x[3] + 
  
  x[1] x[4] + x[2] x[4] + x[3] x[4] + 
   x[1] x[5] + x[2] x[5] + 
   
   x[3] x[5]\}, \{1, -1 - 2 ep, -1 + 
   3 ep\}, 0, \{\{1, 2, 3, 4, 5\}\}]

\normalfont

A similar syntax works for the expansion. In this example the integrator needs access to 
{\tt Mathematica} to evaluate a {\tt PolyGamma} function, so the path is passed
with the {\tt MathematicaBinary} argument:

{\tt SDExpandDirect[var,function,degrees,expand\_var,deltas]}.

\textit{Example}:  

\tt
MathematicaBinary="math";

SDEvaluateDirect[x, \{1, 
  x[1] x[2] x[3] + x[1] x[2] x[4] + 
  
  x[1] x[3] x[4] + x[1] x[2] x[5], 
  x[1] x[3] + x[2] x[3] + 
  
  x[1] x[4] + x[2] x[4] + t (x[3] x[4] + 
   x[1] x[5] + 
   
   x[2] x[5] +  x[3] x[5])\}, \{1, -1 - 2 ep, -1 + 
   3 ep\}, 0, t, 0, 
   
   \{\{1, 2, 3, 4, 5\}\}]

\normalfont

There is one more way to use {\tt FIESTA}. An analytic Feynman integral evaluation method suggested by Lee \cite{Lee:2009dh,Lee:2010cga,Lee:2011jt,Lee:2012te} needs to know
for which values of space-time dimension {\tt d} the integral can have poles. {\tt FIESTA} can locate those values with the use of the following command:

{\tt SDAnalyze[\{U,F,l\},indices,dmin,dmax]}.

Here {\tt dmin} and {\tt dmax} are the ends of the interval where poles should be located. This syntax used only algebraic transformations,
so the result is exact. However, the program might miss some pole cancellations, so some of the returned values might be ``false alerts''.

\textit{Example}:  

{\tt SDAnalyze[UF[\{k\},\{-k$^2$,-(k+p$_1$)$^2$,-(k+p$_1$+p$_2$)$^2$,-(k+p$_1$+p$_2$+p$_4$)$^2$\},
\\
\{p$_1^2\rightarrow$0,p$_2^2\rightarrow$0,p$_4^2\rightarrow$0,
p$_1$ p$_2\rightarrow$-S/2,p$_2$ p$_4\rightarrow$-T/2,p$_1$ p$_4\rightarrow$(S+T)/2,
\\
S$\rightarrow$3,T$\rightarrow1$\}],
\{1,1,1,1\},1,8]
}

Returned answer is {\tt \{2,4\}} which means that the integrand has poles for {\tt d} equal to $2$ and $4$.

\subsection{Program options}

\label{FIESTAoptions}

{\tt FIESTA} has the following options:

\begin{itemize}
 \item {\tt DataPath}: by default {\tt FIESTA} stores databases in the {\tt temp} subfolder. However one might wish to direct it elsewhere, especially if the folder with {\tt FIESTA} is on a network disk. The database should be preferably stored on a fast local disk;
 
 \item {\tt NumberOfSubkernels}: the number of subkernels that {\tt Mathematica} launches. Might be set equal to the number of cores on the computer in use but should not exceed the number of licensed subkernels. This setting can speed up the integrand preparation;
 
 \item {\tt NumberOfLinks}: the number of {\tt CIntegrate} processes that will be launched. The name of this option is left for compatibility with old versions of {\tt FIESTA} where each {\tt CIntegrate} process was called by a separate {\tt MathLink} connection. This setting corresponds to the parallelization during integration;
 
 \item {\tt CubaCores}: the internal parallelization option of the integrators inside {\tt CIntegrate}. By default the integrator uses one core, but it can be changed with this option. This setting also corresponds to the parallelization during integration. Normally {\tt NumberOfLinks} is more efficient, but there might be situations when increasing {\tt CubaCores} leads to better results;
 
 \item {\tt STRATEGY}: sector decomposition strategy. By default we use {\tt STRATEGY\_S} (our strategy), but there are also such options as {\tt STRATEGY\_B} (Bogner and Weinzierl), {\tt STRATEGY\_X} (Binoth and Heinrich), {\tt STRATEGY\_SS} (Speer sectors) and {\tt STRATEGY\_KU}, {\tt STRATEGY\_KU0}, {\tt STRATEGY\_KU2} (Kaneko and Ueda). Among the last three strategies the last variant is the full implementation of the algorithm from \cite{Kaneko:2009qx}, the first two are faster but might result in more sectors;
 
 \item {\tt QHullPath}: if one uses strategies {\tt KU}, {\tt KU0}, {\tt KU2} or the new {\tt SDExpandAsy} syntax, a correct path to the {\tt qhull} executable should be provided. By default it is set to {\tt qhull} assuming that the package is installed on the system, however one might provide a specific path;
 
 \item {\tt CIntegratePath}: by default the integration pool chooses itself the integration binary, however one might provide another path;
 
 \item {\tt UsingC}: by default this option is set to {\tt True}. This means that {\tt FIESTA} uses the {\tt c++} integration. If set to {\tt False}, it switches to {\tt Mathematica} integration, however this is not recommended. With {\tt UsingC} set to {\tt False} the option {\tt ExactIntegrationOrder} (if set) specifies the order in epsilon where {\tt FIESTA} tries exact integration for some time. The default time is $10$ seconds per sector and can be modified by the {\tt ExactIntegrationTimeout} option;
 
 \item {\tt UsingQLink}: by default this option is set to {\tt True}. Switching it off will turn off database usage, however in {\tt FIESTA 3} this is possible only together with {\tt UsingC=False};
 
 \item {\tt OnlyPrepare}: by default this option is set to {\tt False}. In this case the calculation is performed completely, otherwise a database is prepared for integration and a shell command to run {\tt CIntegrate} without {\tt Mathematica} is printed. This can be used if one are preparing a database on one computer and is integrating elsewhere, or if one is willing to try different integrators or precision requests;
 
 \item {\tt SeparateTerms}: if {\tt True}, the integrator receives integrable terms independently, not whole expressions for each sector; on one hand, this simplifies the integrands and the integrators return better precision, on the other hand the error grows after summing up the results, so normally there is no recommendation on whether to use this option or not. However, in the {\tt MPI} mode this option might lead to a significant speedup since in this case it leads to a better parallelization;
 
 \item {\tt ComplexMode} ({\tt False} by default): with this setting set to {\tt True} {\tt FIESTA} performs a contour deformation in order to avoid poles in physical regions. The deformation size depends on a parameter, that is either set manually by giving a value to the {\tt ComplexShift} variable or is tuned automatically in the interval from zero to {\tt MaxComplexShift} ($1$ by default). Increasing the option {\tt LambdaSplit} ($4$ by default) might result in better tuning but slows the preparation; same is true for the search option {\tt LambdaIterations} that is set by default to $1000$;
 
 The contour transformation cannot deal with cases where $F$ turns to zero for the reason that some variables are equal to $1$. In order to trace those cases, set {\tt TestF1=True}. In order to handle those singularities, set the {\tt BisectionVariables} equal to the list of variables such that the integration cube is divided into two parts. By default, the separation point is equal to $1/2$, however this can be changed either by setting the {\tt BisectionPoint} value, or by providing a list of {\tt BisectionPoints};
 
 \item {\tt CurrentIntegrator}: ({\tt vegasCuba} by default): the integrator used at the final stage. The options allowed in the current setup are {\tt vegasCuba}, {\tt suaveCuba}, {\tt divonneCuba} and {\tt cuhreCuba}. It is also possible to add your own integrators by modifying the {\tt integrators.c} source file, however this is far beyond the standard usage of {\tt FIESTA}. Related parameter ({\tt CurrentIntegratorOptions}) presents a list of options of the currently chosen integrator (for details see \cite{Hahn:2004fe}). By default it has no value and the actual options are printed out when one starts the evaluation (the defaults are stored within the {\tt c++} part). The most commonly used integrator option is {\tt maxeval}, which can be set, for example, by {\tt CurrentIntegratorOptions = \{\{"maxeval","500000"\}\}}. The default value is {\tt 50000}. One more virtual integrator is {\tt justEvaluate}. It simply evaluates the integrand at a given point, by default it is the point where all coordinates are equal to $1/2$. Its options are ${\tt x1, x2}$ and so on representing the integration coordinates;
 
 \item {\tt SmallX}, {\tt MPThreshold}, {\tt PrecisionShift}, {\tt MPPrecision}, {\tt MPMinb}: the options that allow to fine-tune the MPFR subsystem. For details see the previous paper on FIESTA \cite{Smirnov:2009pb};
 
 \item {\tt RegVar}: an option introduced for {\tt SDExpandAsy} but usable also in other situations. Sets an extra regularization variable;
 
 \item {\tt AnalyticIntegration}: an option used only for {\tt SDExpandAsy}, {\tt True} by default, tells {\tt FIESTA} to try to take some integrations analytically after introducing regions;
 
 \item {\tt FastASY}: ({\tt False} by default) specifies the region search mode (used in {\tt SDExpandAsy}). With {\tt FastASY} set to {\tt False} the polynomial {\tt U $\times$ F} is analyzed, with {\tt True} --- the $F$ polynomial. The {\tt FastASY} variant might work significantly faster and will produce correct results almost all the time, but one should use it at his own risk;
 
 \item {\tt PolesMultiplicity}: an option used only for {\tt SDAnalyze}, {\tt False} by default, changes the answer so that it returns not only values of {\tt d} but also maximal pole multiplicities;
 
 \item {\tt {\tt Mathematica}Binary}: a path to the executable {\tt Mathematica} kernel. If set, it is passed to the integration pool, so it can request {\tt Mathematica} for values of functions it cannot evaluate itself (currently this feature is used only for {\tt PolyGamma});
 
 \item {\tt BucketSize} ($25$ by default): an option tuning the database usage (for details see the documentation on KyotoCabinet); increasing this variable might result is faster database access, but increases the RAM usage;
 
 \item {\tt MixSectors} ($0$ by default): lets {\tt FIESTA} to sum up integrands in different sectors before integration; 
 
 \item {\tt RemoveDatabases} ({\tt True} by default): specifies whether the database files should be removed after the integration;
 
 \item {\tt d0} ($4$ by default): specifies the space-time dimension;
 
 \item {\tt ReturnErrorWithBrackets}: ({\tt False} by default) changes the output --- with {\tt True} the error estimates are printed as {\tt pm[NUMBER]} instead of {\tt pmNUMBER};

 \item {\tt FixSectors}: ({\tt True} by default) --- for the reason of fixing sector numbers and easier debugging we perform the variable substitutions stage on the main kernel. If one sets this option to false, this stage will be also made parallel, however it normally does not influence the total time much;
 
 \item {\tt PrimarySectorCoefficients}: one might specify the list of coefficients before primary sectors. A zero means that this sector will be ignored. With this setting one can split the problem into parts and also take diagram symmetries into account;
 
 \item {\tt NoDatabaseLock}: prevents {\tt FIESTA} from locking the database. This may be away to avoid restrictions on some file systems but might result in corrupted databases.
 
\end{itemize}

\subsection{CIntegratePool options}

\label{PoolOptions}

If one runs {\tt FIESTA} with {\tt OnlyPrepare=True}, then it prints out the command to be executed, for example,

\noindent {\tt bin/CIntegratePool -in /temp/db2in -out /temp/db2out}

\noindent {\tt -all -direct -threads 4 -complex}

After executing such a command and achieving a result, one might wish to rerun the integration with different options. 
Here we provide the list of arguments accepted by {\tt CIntegratePool}:

\begin{itemize}
 \item {\tt -in}: provides the path to the database with integrands (without the .kch suffix);
 \item {\tt -out}: provides the path to the database where results are stored;
 \item {\tt -direct}: instructs {\tt CIntegratePool} that it was called directly and not from {\tt Mathematica}, so that it saves the results in the output database, does not use temporary files in order to transfer results back and prints the results to {\tt stdout};
 \item {\tt -math}: provides the path to the {\tt Mathematica} binary;
 \item {\tt -bucket}: provides the bucket value for the output database;
 \item {\tt -CIntegratePath}: provides the path to the {\tt CIntegrate} binary. This can be either a full path (starting with {\tt /}) or just a filename, in this case {\tt CIntegratePool} searches for this file in the same directory. If this option is missing, it searches either for {\tt CIntegrateMP} or {\tt CIntegrateMPC} depending on whether the complex mode is on or off;
 \item {\tt -integrator}: sets the integrator to be used. {\tt -intpar} sets some integrator parameter, for example, {\tt -intpar maxeval 500000}. For the list of integrators see section~\ref{FIESTAoptions};
 \item {\tt -MPThreshold}, {\tt -MPPrecision}, {\tt -PrecisionShift}, {\tt -SmallX}, {\tt -MPMin}: options that are fine-tuning the {\tt MPFR} subsystem;
 \item {\tt -threads}: sets the number of {\tt CIntegrate} processes launched by the pool. This option is meaningless in the {\tt MPI} mode;
 \item {\tt -CubaCores}: sets the number of processes that the integrator starts for each integrand;
 \item {\tt -test}: perform an integrator test only, {\tt -notest}: do not perform this test, {\tt -nopreparse}: do not perform parse check of all expressions before integration;
 \item {\tt -complex}: specifies that the expression is complex. If one knows it do be real, this setting should not be used since it can slow down the integration a lot;
 \item {\tt -all}: perform all integration; to the contrary, {\tt -task} followed by a number instruct the code to evaluate only expression related to one {\tt SDEvaluate} call. Normally, there is only one task in the database, so one would call {\tt -task 1}, but the {\tt SDExpandAsy} mode uses multiple tasks. {\tt -prefix} can be used only when {\tt -task} is set and tell the program to integrate only with given powers of $\ep$ and {\tt RegVar}. For example {\tt -task 1 -prefix "\{-2,-\{1, 2\}\}"} corresponds to integrals having $\ep$ order $-2$ and {\tt RegVar} coefficient {\tt RegVar * Ln [RegVar]\^{ }2};
 \item {\tt -separate\_terms}: instructs the algorithm not to group expressions by sectors and to integrate each integrable term separately, might be useful if one is using massive {\tt MPI} parallelization;
 \item {\tt -testF}: instead of the integration, the code checks whether the sign of the imaginary part of $F$ is negative. Might be used for debugging special cases in complex mode, for details see section \ref{complex};
 \item {\tt -debug}: used to print all integration results.
\end{itemize}

\subsection{CIntegrate options}
The integration pool program, {\tt CIntegratePool} or {\tt CIntegratePoolMPI} distributes tasks between on of the integration programs provided with the package --- {\tt CIntegrate}, {\tt CIntegrateMP}, {\tt CIntegrateMPC} or even something external. However, the integration programs can be called on their own. They have no options on start, accept input from {\tt stdin} and print it to {\tt stdout}. Each command sent to the program is ended with a new line symbol.

The main command to be provided to the program input is {\tt Integrate}. After that one should send the expression. It consists of a number of lines, each of them should be ended with the {\tt ;} symbol. At the end should be a line consisting of the {\tt |} symbol. The expression lines are the following:

\begin{itemize}
 \item The number of variables;
 \item The number of intermediate functions;
 \item A number of lines each representing an intermediate expression. If the second line is {\tt 0;}, then this part should be missing;
 \item The final expression.
\end{itemize}

The expressions might contain algebraic operations such as {\tt +, -, *, /}, bracket symbols, numbers with a floating point. The integration variables should be referred as {\tt x[1], x[2]} and so on, intermediate functions as {\tt f[1], f[2]} and so on. Power is represented as {\tt p[expr,exponent]}, natural logarithm as {\tt l[expr]}. One can also use {\tt P} for $\pi$ and {\tt G} for {\tt EulerGamma}. {\tt PolyGamma[arg1,arg2]} also works but one needs to provide a path to the {\tt Mathematica} binary --- the integrator cannot evaluate this function on its own.

\textit{Example}:  

\tt

1;

0;

x[0]+0.2;

|

\normalfont

If one feeds this example into the {\tt CIntegrate} program after the {\tt Integrate} command then it will result in integrating $x+0.2$ from $0$ to $1$.

The program also accepts the following commands (most of them require an argument passed as next line):

\begin{itemize}
 \item {\tt Parse}: same as {\tt Integrate}, but the expression is only parsed;
 \item {\tt Exit}: quit the program;
 \item {\tt CubaCores}: sets the number of cores used by the integrator, default value is $1$;
 \item {\tt SetMath}: provides a path to the {\tt Mathematica} binary;
 \item {\tt SetIntegrator}: sets the integrator to be used;
 \item {\tt SetCurrentIntegratorParameter}: sets one of the integrator parameters, the next line should be the parameter name, the line after that --- the value;
 \item {\tt GetCurrentIntegratorParameters}: simply returns the list of current parameters and their values;
 \item {\tt MPFR}, {\tt Native} and {\tt Mixed} (default variant): chooses whether the integrand should use {\tt MPFR} everywhere, the double precision or the mixed mode. The mixed mode used the following five options to determine in which parts of the integration cube which arithmetics should be used: {\tt SetMPPrecision}, {\tt SetMPPrecisionShift}, {\tt SetMPMin}, {\tt SetMPThreshold}, {\tt SetSmallX};
 \item {\tt Debug}: makes the code print values in all integration points;
 \item {\tt TestF}: instead of the integration the code checks the sign of the imaginary part of the integrand;
 \item {\tt Help}: list all those commands.
\end{itemize}

\subsection{Setup for cluster calculations}

In order to perform integrations on a cluster one should do the following:

\begin{enumerate}
 \item Run {\tt FIESTA} with {\tt OnlyPrepare=True}.
 \item Save the database file produced by the integrand preparation stage.
 \item Use the command printed out by the {\tt Mathematica} part to launch it on a cluster. One can replace the call to {\tt CIntegratePool} with {\tt CIntegratePoolMPI} in order to use the {\tt MPI} version. The syntax to launch {\tt MPI} program varies, so one will have to use instructions for the cluster in use. One can also adjust the integrator options, for example by increasing the number of sampling points.
 \item The result is printed out in the {\tt c++} log and also saved to a small output database. To see the result in the {\tt Mathematica} form one should load {\tt FIESTA}, provide the proper {\tt DataPath} and run {\tt GenerateAnswer[]}. No more options are required.
\end{enumerate}

\section{Conclusion}

We have presented the new version of {\tt FIESTA} --- a program for automatic numerical evaluation and analytic expansion of Feynman integrals.
The new version contains new sector decomposition and expansion algorithms, provides possibilities to integrate in physical regions 
and to perform cluster parallelization. We believe that this upgrade is an essential improvement in automatic numerical evaluations
of Feynman integrals.
{\tt FIESTA} development is not over. One of the future plans is to make use of {\tt GPU}s in order to speed up the integration.

\section*{Acknowledgements}

This work was supported by DFG through SFB/TR 9.
I would like to thank P.~Marquard for numerous tests of the beta version of {\tt FIESTA},
M.~Tentyukov for advices on code optimization 
and V.~Smirnov and M.Steinhauser for ongoing support as well as the careful reading of the draft of this paper.

\bibliographystyle{model1-num-names}
\bibliography{FIESTA3,asmirnov}
\end{document}